 \documentclass[final,3p,times]{elsarticle}

 \usepackage{graphicx}

\usepackage{amssymb}
  \usepackage{amsthm}

\newcommand{\PF}{(TMTSF)$_2$PF$_6$}
\newcommand{\As}{(TMTSF)$_2$AsF$_6$}
\newcommand{\X}{(TMTSF)$_2$$X$}
\newcommand{\Ts}{ $T_{\rm \small SDW }$} 

\newcommand{\Ti}{  $T_1^{-1}$}
 \newcommand{\TT}{  $(T_1T)^{-1}$}
 \newcommand{\bk}{ \bar{k}}
\journal{Comptes Rendus Physique}

\begin{document}

\begin{frontmatter}


\title{Superconductivity and antiferromagnetism as interfering orders in organic conductors}


\author{C. Bourbonnais$^{1,2}$ and A. Sedeki$^{1}$}

\address{$^1$Regroupement Qu\'eb\'ecois sur les Mat\'eriaux de Pointe, D\'epartement de Physique, RQMP, Universit\'e de Sherbrooke, Sherbrooke, Qu\'ebec, Canada, J1K-2R1}
\address{$^2$ Canadian Institute of Advanced Research, Toronto, Canada. }
\begin{abstract}
Superconductivity in the Bechgaard salts series of quasi-one-dimensional organic conductors occurs on the verge of spin-density-wave ordering  when  hydrostatic pressure is applied. The sequence of instabilities is intimately connected to  normal state anomalies in various quantities like the temperature dependence of electrical transport and nuclear spin-lattice relaxation rate. We discuss how such a connection  takes  its origin in the  interference between the different pairing mechanisms   responsible for antiferromagnetism and superconductivity,  a duo that can be comprehended in terms of a weak  coupling renormalization group theory. The recent developments along this line of though are presented in relation to experiments.  
\end{abstract}


\end{frontmatter}


\section{Introduction}
\label{Intro}
Superconductivity   in organic conductors  was first   observed  more than three decades ago  in the quasi-one-dimensional metal (TMTSF)$_2$PF$_6$ (bi-tetramethyltetraselenafulvalene hexafluorophosphate),  a member of  the famous Bechgaard salt series,   (TMTSF)$_2X$ ($X$= PF$_6$, AsF$_6$, ...) \cite{Jerome80}. Manifestation of  superconductivity (SC) in this series   emerges under pressure close by a magnetic  instability of the metallic state against   the formation of  spin-density-wave (SDW) order  \cite{Jerome82,Bourbon08}. This sequence of phases along with its share of anomalies in the normal state,    soon appeared rather unfamiliar and at odds with what was  found until then in  classical superconductors. The closeness of antiferromagnetism and superconductivity  was  not    unique to the Bechgaard salts, but was  also present  in  different classes of materials among which the little older `heavy fermions' materials \cite{Steglich79,Mathur98},  followed  later on by  the  high-$T_c$ cuprates \cite{Bednorz86,Taillefer10,Armitage10}, layered organics \cite{Laukhin85,Urayama88,Kanoda97,Lefebvre00,Powell06}, and more  recently by   iron-based superconductors \cite{Kamihara08,Fang09}. 

In spite of  the considerable progress achieved in the understanding of  correlated electron systems in the last decades, no consensus is yet to be found about the role played by antiferromagnetism for Cooper pairing  in each of the above categories of materials, as well as about  the possibility to extract some common view  that would link   pairing with magnetism across them.  In face of these difficulties, the study of low-$T_c$ superconductors like the Bechgaard salts  may represent  a simpler avenue of investigation,  providing key insight into this  enduring issue. Experiments in these molecular compounds are   generally compatible with  the pronounced itinerant character of carriers and the existence of a  well-defined  quasi-one-dimensional -- open -- Fermi surface. This provides substantial grounds for a   relatively simple weak coupling  formulation of  the problem of competing  instabilities taking place in   these anisotropic electron systems. Its best formulation  is given   in terms of the  quasi-1D electron gas model whose consequences  can be worked out fairly well by the renormalization group (RG) method \cite{Bourbon09}.     RG proved to be a successful strategy  for taking into account  the contribution of  both particle-hole  and particle-particle  scatterings responsible for SDW and SC pairings in the vicinity of the Fermi surface. Appearing  simultaneously in perturbation theory, these scattering processes interfere with  one another at arbitrary order,  which can lead    for instance to their  mutual reinforcement. The consequences of this mixing are far from being trivial.  Under certain conditions  related to  nesting deteriorations in  the particle-hole channel,   resulting for example from  the application of pressure,  interference  acts   as the  mechanism  of onset of magnetic  pairing for  d-wave superconductivity from an inhibited  SDW instability. The SDW-SCd sequence of transitions thus obtained versus nesting alterations has been found to  capture    the essentials of the phase diagram of the Bechgaard salts under pressure.

The impact of interference between     primary scattering channels  is  not restricted to the transition lines of the phase diagram, but is found to   affect   the properties of the normal phase as well. It is where  it can supply    a consistent explanation  for the  observation of a linear$-T$   resistivity \cite{DoironLeyraud09,DoironLeyraud10},  and  a Curie-Weiss nuclear relaxation rate \cite{Wu05,Brown08}, whose  amplitudes show some amazing correlation with the superconducting instability line of the phase diagram under pressure.

The purpose of this capsule review is to discuss  some recent experimental results about the  phase diagram   of the prototype organic superconductor  (TMTSF)$_2$PF$_6$, alongside electrical resistivity and NMR relaxation rate anomalies of its    metallic state. We then summarize  the latest theoretical developments of the  renormalisation group method in      the framework  of the quasi-one-dimensional electron gas model,  providing the logical link needed to  understand these key features displayed by  quasi-one-dimensional  organic superconductors.

\section{ Bechgaard salts: some background of experimental results}

Although early band calculations predicted that these systems should be metallic, the first ambient pressure resistivity measurements on the prototype  compounds  of the series,  \PF\  and \As,  revealed instead the existence of  a metal-insulator transition at $T_{\rm SDW}\sim 12$~K  \cite{Bechgaard80}. Contrary to the situation encountered in organic materials that preceded the discovery of the Bechgaard salts, like  the celebrated compound TTF-TCNQ \cite{Jerome82}, the low pressure state was not the result of a charge-density-wave or a Peierls instability, but was rather  magnetic in character caused by the onset of  SDW order. The fingerprints  of long-range SDW correlations at the approach of $T_{\rm SDW}$ have been evidenced by several techniques. The  Nuclear Magnetic Resonance (NMR) probe is one of them.  At the approach of $T_{\rm SDW}$, critical SDW fluctuations give rise to a singular growth of   staggered local magnetic field   resulting in    a  square-root singularity of the NMR  spin-lattice relaxation rate, $T_1^{-1}$,  as shown by the $^{77}$Se NMR data  of Figure~\ref{T1}-b for \PF\  \cite{Brown08,Creuzet87b}. Furthermore, from the  analysis of the proton  ($^1$H) NMR line shape in the SDW ordered state of  \PF, it was also possible to simulate the spatial distribution of the dipolar field and  then extract  the SDW modulation wave vector ${\bf q}_0$ \cite{Takahashi86}.    The value ${\bf q}_0=(2k_F,{1\over 4}b^*)$ thus obtained for the staggered magnetization  in the $ab$ plane  coincides,  within experimental accuracy,  with the best nesting vector of the Fermi surface found by  band calculations \cite{Ducasse86}. This correspondence is of importance since it   shows that the SDW transition  in \X \ is  an  instability of the electron system driven,  besides  repulsive interactions,  by  nesting of the Fermi surface at the same  ${\bf q}_0$, an important factor in favor  of a weak coupling description of the SDW instability in these materials.
 
\begin{figure}
 \centerline{\includegraphics[width=15.0cm]{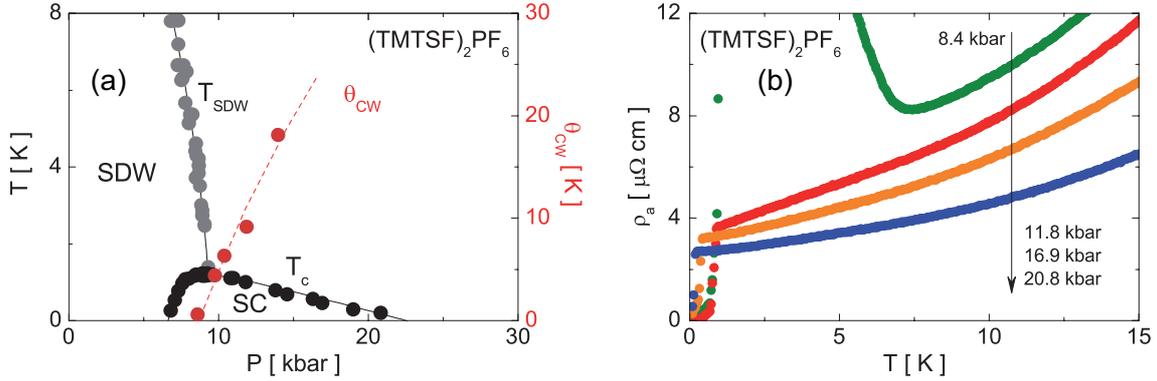}}
   \caption{a) $P$ressure$- T$emperature phase diagram of  \PF\ (after Ref.~\cite{DoironLeyraud09}); also shown the variation of the Curie-Weiss scale $\Theta$ under pressure extracted from the NMR data of Fig.~\ref{T1}-b \cite{Brown08} (owing to a different low temperature determination  of pressure in these two sets of experiments,  a rigid, positive, shift of 3 kbar has been applied to the pressure scale  of  Fig.~\ref{T1}-b);  b) Longitudinal resistivity {\it vs} temperature for \PF\ at different pressures. \label{Phases}} 
   \end{figure}

\begin{figure}
  \centerline{\includegraphics[width=15.0cm]{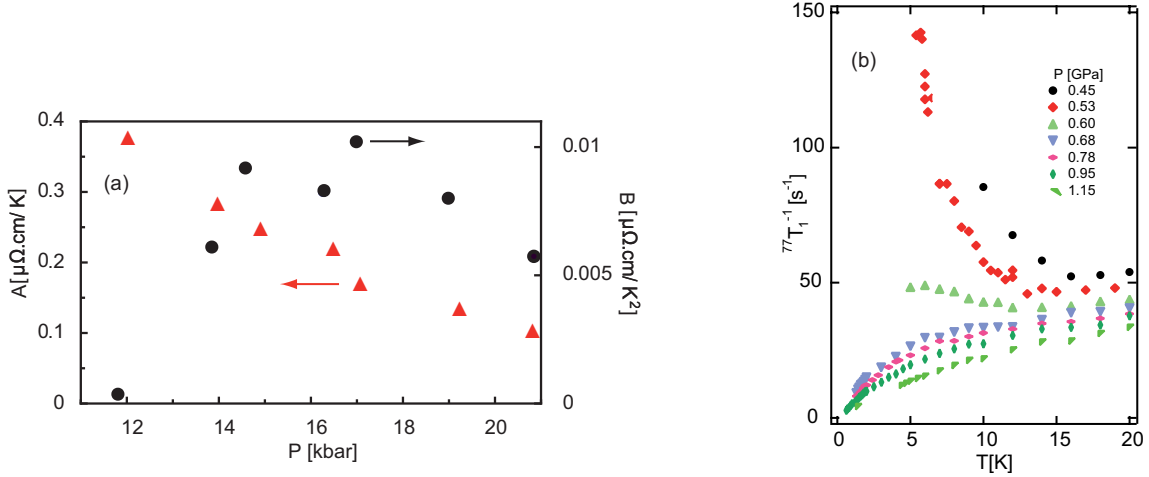}}
\caption{(a) Pressure dependence of the resistivity coefficients $A$ and $B$ of the polynomial fit $\rho_a(T) =   \rho_0 + AT + B T^2$ of the data of Figure~\ref{Phases}-b; after Ref.~\cite{DoironLeyraud10}. (b) Temperature dependence of  spin-lattice nuclear relaxation rate \Ti in \PF\ at different pressures; after ref.~\cite{Brown08}. \label{T1}} 
   \end{figure}

 Under hydrostatic  pressure, \Ts \  decreases smoothly  and ultimately undergoes  a rapid decline near a pressure threshold    of about 8~kbar for \PF. There, the resistivity first shows an insulating SDW behavior and then drops to zero following  the onset of superconducting order at the temperature $T_c$ (Fig. \ref{Phases}-a). Reentrant  superconductivity within the SDW phase is apparent over an  interval of more or less one kbar in which  SC   coexists  in a segregated form with SDW   \cite{Vuletic02,Lee02b}. By increasing pressure up to some critical value  $P_c$,  $T_c$ reaches a maximum $(\sim  1.2$~K)  that  defines the SDW-SC juncture  where \hbox{$T_{\rm SDW} \approx T_c$}. Above $P_c$, reentrance   gives way to a simple metal to SC transition characterized by   a rapid  downturn of $T_c$     toward very small values beyond  20 kbar  (Fig.~\ref{Phases}-a).   The nature of the superconducting order parameter in (TMTSF)$_2X$,   either  singlet or triplet,  along with    the presence of  nodes or not   on the Fermi surface, have been  much debated. However, the current experimental status appears to  tip the scale in  favor of a singlet order parameter with nodes \cite{Shinagawa07,Joo05,Yonezawa11}.  
 
 A striking feature of  the normal phase  resistivity,   close to $P_c$, is shown by its temperature  dependence: the longitudinal resistivity $\rho_a$ accurately   fits the non Fermi liquid linear expression, $\rho_a(T)=\rho_{0,a} + AT$, where $\rho_{0,a}$ is the residual resistivity and $A>0$ \cite{DoironLeyraud09,DoironLeyraud10}.  In \PF\ at   11.8 kbar for example,  linear--$T$ resistivity is detected from 8~K or so  down to $T_c$; it  even extends toward  much lower temperature  when a small magnetic field is applied to suppress $T_c$. Away from $P_c$, $\rho_a$ {\it vs} $T$ modifies in a distinctive  way. Resistivity ceases to be exclusively linear and  acquires some curvature  that satisfactorily   fits   the polynomial form $\rho_a(T)=\rho_{0,a} + AT +BT^2$. This can be interpreted as  a  Fermi liquid component  that sets  in under pressure, becoming in turn the main contribution to resistivity    at very high pressure, namely where the linear$-T$ resistivity component, like $T_c$,     becomes vanishingly small   (Figs.~\ref{Phases}-b and \ref{T1}-a).   This notable correlation between $A$ and the strength of superconductivity   shows that inelastic scattering and Cooper pairing are interrelated. This feature turns out to be not unique to the Bechgaard salts \cite{DoironLeyraud09,DoironLeyraud10,Taillefer10},  but   is also shared by other categories of unconventional superconductors showing close proximity with antiferromagnetic ordering, especially in the pnictides \cite{DoironLeyraud09,Kasahara10} and  high-Tc cuprates \cite{MacKenzie96,AbdelJawad06,Taillefer10,Taillefer06}.   
 
 Connections between superconductivity and deviations from the Fermi liquid predictions in the same region of the normal state are not restricted to electrical transport, but are also found in  other quantities. The  NMR spin-lattice relaxation rate  \Ti\ is one of these \cite{Brown08,Creuzet87b}. If one first looks at the magnetic sector of the phase diagram of Figure~\ref{Phases}-a,   \Ti\ of Figure~\ref{T1}-b displays the characteristic square root singularity $\sim (T-T_{\rm SDW})^{-1/2}$,   which confirms the onset of long-range three-dimensional SDW correlations at the approach of \Ts \ \cite{Creuzet87b,Wzietek93}. In the superconducting sector above $P_c$, the critical behavior is suppressed, but  an anomalously large enhancement of \Ti\ remains (Figure~\ref{T1}-b);  its   amplitude  is huge close to $P_c$ and  reduces progressively as pressure is raised. The origin of the anomaly lies in the presence of short-range SDW spin fluctuations whose amplitude shows   persistent growing  down to $T_c$ \cite{Bourbon84,Bourbon09}, that is in the same temperature region   where  linear$-T$ resistivity is seen (Fig.~\ref{Phases}-b). The temperature profile of   relaxation rate  differs there from    the Korringa law,   \TT  $\sim {\rm const.}$,   expected for a Fermi liquid; it rather exhibits an enhancement following the Curie-Weiss behavior $(T_1T)^{-1} \sim (T+\Theta)^{-1}$ \cite{Wu05}. The Curie-Weiss scale, $\Theta$,  linked to the amplitude of the anomaly and in a way to the characteristic energy scale of spin fluctuations, is rapidly  changing under pressure: close to the SDW-SC juncture, $\Theta$ is vanishingly small, indicative of  its  critical suppression as $ P\approx P_c$. From $P_c$ upward, $\Theta$  raises rapidly  and reaches large values at high pressure (Fig.~\ref{Phases}-a).
 
 The  anomaly in the temperature dependence of  the nuclear relaxation rate,  along with its modification under pressure hints at a direct participation  of spin fluctuations in linear$-T$ resistivity above $T_c$.  Low-energy SDW fluctuations   evidenced  by the low-frequency NMR probe can  act as an important source of scattering for electrons  and can then  influence the   resistivity in a significant way.  At the same time, the same SDW fluctuations   have the ability to   promote unconventional   d-wave  Cooper pairing.  This may  explain the amazing   correlation between the  non Fermi liquid features of the normal state  and superconductivity whose importance is best meant by the size of $T_c$.

\section{Theory of the quasi-one-dimensional electron gas model}
\subsection{Model}
Despite the apparent molecular complexity of \X \  solids, their low energy electronic structure can be approached in a relatively simple way with the aid of molecular orbitals. Dominant van der Waals  bonding yields a strongly anisotropic overlap of   the  highest occupied  $\pi$ molecular orbitals, the so-called HOMO's of the TMTSF molecule. The HOMO is partly filled owing to the charge transfer of one electron  per  two molecules  towards the radical $X$, as indicated by the 2:1 stoichiometry of the compound. An extended-H\"uckel analysis of the electronic structure reveals the presence of a single band  crossing the Fermi level \cite{Grant83,Ducasse86}. A  simple one-component quasi-1D open Fermi surface   follows, which consists of two warped planes centered around  the one-dimensional Fermi wave vector $\pm k_F$ of  isolated chains. The TMTSF stacks, however, are not uniform  and display  a slight dimerisation which doubles the longitudinal lattice periodicity and then opens a small gap, $\Delta_D$, at $\pm 2k_F$, namely  in the middle of the 3/4-filled band of  the uniform limit. There is one hole per dimer in average   and the band can be considered as effectively half-filled at low energy or temperature compared to $\Delta_D$ \cite{Emery82}. An important outcome of the      half-filled character of the band, though weak,   is the possibility of electron-electron Umklapp scattering along the stacks. This   reinforces  the impact of nesting, which is at the core of instabilities of the electron system toward density-wave formation.

A suitable tight-binding quasi-1D model of the  energy spectrum  is given by 
\begin{equation}
\label{Ek}
E_p(\mathbf{k}) = v_F(pk-k_F) -2t_{\perp b}\cos k_b -2t_{\perp b}'\cos 2k_b - 2t_{\perp c}\cos k_c,
\end{equation}
for which the longitudinal tight-binding part  $-2t_a\cos k$  has been linearized around the 1D Fermi points $\pm k_F$ $(\pm \pi/2)$;  here $v_F=2t_a$ ($E_F= v_F k_F$) corresponds to the longitudinal Fermi velocity (energy) and $p$ refers  to right ($p=+$) and left ($p=-$) moving carriers. The anisotropy in the band structure  is about  $E_F\sim 15t_{\perp b}\sim 450 t_{\perp c}$, along the $a$, $b$ and $c$ directions, which are  taken    orthogonal for simplicity \cite{Grant83,Ducasse86}. A next-to-nearest neighbor hopping term of  amplitude $t_{\perp b}' \ll t_{\perp b}$     along the transverse $b$ direction enters in the definition of $E_p(\mathbf{k})$. This term violates the electron-hole symmetry of the   spectrum (\ref{Ek}), i.e., $E_p(\mathbf{k} +\mathbf{q}_0)\ne-E_{-p}(\mathbf{k})$ at the  nesting wave vector $\mathbf{q}_0=(2k_F,\pi, \pi)$. Deviations are introduced by  a finite $t_{\perp b}'$, which  simulates  the main influence of pressure on  nesting.

The partition function of   the quasi-1D electron gas model, $Z=\int\!\!\int \mathfrak{D}\psi^* \mathfrak{D}\psi \ e^{S_0+ S_I}$,  when expressed as a  functional integral
      over  fermion fields,  is parametrized  by an action $S$  whose    quadratic part, $S_{\!0}$, for free electrons is given by
\begin{equation}
\label{S0}
S_{\!0}[\psi^*,\psi] =\sum_{p,\sigma,\bk} [G^0_p(\bk)]^{-1} \psi^*_{p,\sigma}(\bk)\psi_{p,\sigma}(\bk),
\end{equation}
where 
\begin{equation}
\label{G0}
G^0_p(\bk) = [i\omega_n-E_p(\mathbf{k})]^{-1},
\end{equation}
is the bare fermion propagator and $\bk=(\mathbf{k},\omega_n=\pm \pi T,\pm 3 \pi T, \ldots)$.  The interacting part $S_I$ takes the form
$$
S_I   =   -   {1\over TLN_\perp} \sum_{\{\bar{k},\sigma\}} \,  \big[\, g_{\{\sigma\}}(k'_{b1},k_{b1},k_{b2}) \, \psi^*_{+,\sigma'_1}( \bar{k}'_1)\psi^*_{-,\sigma'_2}( \bar{k}'_2)\psi_{-,\sigma_2}( \bar{k}_2)\psi_{+,\sigma_1}( \bar{k}_1)  
$$
\begin{equation}
\label{SI}
  + \ {1\over 2} g_{3}(k'_{b1},k_{b1},k_{b2}) \,\big( \psi^*_{+,\sigma_1}( \bar{k}'_1)\psi^*_{+,\sigma_2}( \bar{k}'_2)\psi_{-,\sigma_2}( \bar{k}_2)\psi_{-,\sigma_1}( \bar{k}_1)+ {\rm c. c.} \, \big)\big]\delta_{ \bar{k}_1+ \bar{k}_2= \bar{k}'_1+ \bar{k}'_2(\pm \bar{G}) }. 
\end{equation}
In the   g-ology   prescription, the   electron-electron interaction separates into   normal and Umklapp processes. The  former part,  
 \begin{equation}
\label{ }
g_{\{\sigma\}}(k'_{b1},k_{b1},k_{b2}) = g_2\delta_{\sigma_2\sigma'_2}\delta_{\sigma_1\sigma'_1} -g_1\delta_{\sigma'_1\sigma_2}\delta_{\sigma'_2\sigma_1},
\end{equation} 
  retains the backward ($g_1$) and forward ($g_2$) bare scattering amplitudes between right and left moving carriers. As for Umklapp scattering, for which $\bar{G}=(4k_F,\bf{0})$, it is defined by the bare amplitude $g_3(k'_{b1}, k_{b1},k_{b2})= g_1\Delta_D/E_F$, which is small  for  weakly dimerized chains. This is the situation that prevails in the Bechgaard salts where   one has typically $\Delta_D/E_F \lesssim  0.1$ \cite{Emery82}.  All the above scattering amplitudes    are independent of the momentum at the bare level.  They  are commonly expressed in terms of the one-site and nearest-neighbor site couplings parameters $U$ and $V$ of the extended Hubbard model in the continuum (electron gas) limit, namely $g_1=U-2V$, $g_2=U+2V$, and $g_3= g_1\Delta_D/E_F$.

The range of various band and  coupling parameters of the above quasi-1D electron gas model can be fixed from various sources. Band calculations, as well  experiments are compatible with the following typical set of values for the kinetic part of the Hamiltonian,  $E_F\simeq 3000$~K, $t_{\perp b}\simeq 200$~K and $t_{\perp c} \lesssim 10$~K. As for the bare amplitude of the $g_i$$\,{'s}$, the observed enhancement of uniform magnetic susceptibility  can be called for to fix the  range of the  backscattering amplitude $g_1$ \cite{Wzietek93}. The analysis of the temperature dependence of  susceptibility is consonant with a value revolving around $\tilde{g}_1\equiv g_1/\pi v_F\sim 0.3$ (henceforth  normalized  by the longitudinal bandwidth). Given the  size of the dimerization gap $\Delta_D$, this fixes  the amplitude of Umklapp scattering at a small, but finite value,  $\tilde{g}_3\simeq 0.02$. Finally, the bare $\tilde{g}_2$ can be estimated by   the value needed to match the   optimal \Ts\  observed to the calculated scale, as obtained from RG with the above set of figures in the limit of small $t_{\perp b}'$. Thus the maximum   $T^0_{\rm SDW}$\ ($\sim 25$K) found on experimental grounds in the cousin compounds (TMTTF)$_2X$   \cite{Klemme95},  yields $\tilde{g}_2\sim  0.6$.

\subsection{One-loop renormalisation group results} 
Long-distance correlations and the  propensity for ordering   in the most conducting  -- $ab$ -- plane can be studied  by the RG method \cite{Duprat01,Nickel06,Bourbon09}. In this approach, one proceeds in the partition function $Z$ to the successive integration  of electronic degrees of freedom $\psi^{(*)}$, from the high energy cutoff $E_F$ down to the energy distance $E_Fe^{-\ell}$  above and below each Fermi sheet at step $\ell\ge0$. Constant energy surfaces at  $\ell$  are divided into a number patches \cite{Zanchi00,Nickel06}, each being  indexed by  a particular  $k_b$ value for the momentum  along   $b$. 
  
\begin{figure}
 \centerline{ \includegraphics[width=7.0cm]{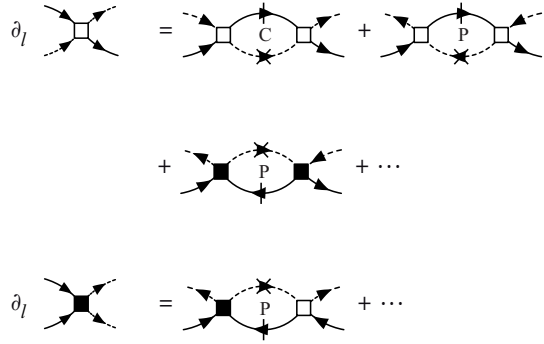}}
\caption{One loop RG flow equations for the normal $g_{1,2,}$ (open square) and Umklapp $g_{3}$ (full square) scattering amplitudes. Here $C$ and $P$ stand for the Cooper  and Peierls loops. The full (dashed) lines refer to right (left) moving electrons.\label{1loop}} 
   \end{figure}

  At the one-loop level, the partial integration leads to successive corrections to the scattering amplitudes $\tilde{g}_i$ as function of $\ell$ and  for a given temperature $T$. These 
 come  from the logarithmically singular loops of particle-particle (Cooper)
and particle-hole (Peierls) scattering channels. Both contributions generate momentum dependence for the couplings, which is retained for the $b$ direction only. This leads to the 
  flow equations of Fig.~\ref{1loop} \cite{Nickel06,Duprat01}, which are written in the  schematic form 
  \begin{eqnarray}
\label{gn}
\partial_\ell \tilde{g}_{i=1,2}(k'_{b1},k_{b2},k_{b1}) & = & \sum_{n,n'=1}^3\Big\{\epsilon^{n,n'}_{C,i}\,\left\langle \tilde{g}_n  \cdot \tilde{g}_{n'} \cdot \partial_\ell {\cal L}_C\right\rangle_{k_b}  
+  \epsilon^{n,n'}_{P,i}\,\left\langle   \tilde{g}_n  \cdot \tilde{g}_{n'} \cdot \partial_\ell{\cal L}'_P\right\rangle_{k_b} \Big\}, \cr
\partial_\ell \tilde{g}_3(k'_{b1},k_{b2},k_{b1})   &  = &  \sum_{n =1}^2 \epsilon^{3,n}_{P,3} \,\left\langle \tilde{g}_3 \cdot \tilde{g}_{n}\cdot \partial_\ell{\cal L}_P\right\rangle_{k_b}, 
\end{eqnarray}
where the momentum dependence of various products has been masked  for simplicity. Here the Cooper and Peierls loops ${\cal L}_P(k_b, q_{P},t'_{ b})=T\sum_{\omega_n} \int dE_P[G^0_p(\bar{k})]_\ell [G^0_{-p}(\bar{k}-\bar{q}_P)]_\times$ and ${\cal L}_C(k_b, q_{C})=T\sum_{\omega_n}\int  dE_P$ $[G^0_p(\bar{k})]_\ell [G^0_{-p}(-\bar{k} +\bar{q}_C)]_\times$ are  evaluated at $q_P = k_{b1}-k_{b1}'$ and $q_C=k_{b1}+k_{b2}$, respectively. These are expressed as integrals over products of propagators  on ($[G^0_{-p}]_\ell$) and above $([G^0_{-p}]_\times)$ the energy shell at  $E_F e^{-\ell}$ . Here   $\langle \ldots \rangle_{k_b}$ stands for an average over $k_b$ and  $\epsilon_{C,P,i}^{nn'}$ are   the coefficients that fix  the sign of  closed loops ($\epsilon_{P,i}^{nn'}=-2)$, vertex corrections ($\epsilon_{P,i}^{nn'}=1)$ and ladder graphs ($\epsilon_{C,i}^{nn'}=-\epsilon_{P,i}^{nn'}=-1$). These different contributions will then interfere with one another as a function of $\ell$ or the   energy distance $E_Fe^{-\ell}$ from the Fermi surface. For  small $\Delta_D$ and then weak initial Umklapp term,  interference for high energy -- 1D -- degrees of freedom where $E_Fe^{-\ell} > t_b$,   give rise to   coupling constant flows   differing  very little from those obtained in  the   1D    (Luttinger liquid) limit.  As $E_Fe^{-\ell} < t_b$, however, sensitivity to the  wrapping  of the Fermi surface develops  and interference modifies accordingly, becoming  non uniform along the Fermi surface. The $g's$ then acquire a $k_b$ dependence and their flows can be governed by a strong coupling fixed point where the $g's$ become singular, signaling an instability of the metallic state against   ordering.  

For the above model of a quasi-1D metal with  repulsive intrachain interactions and weak dimerization, long-range order can only occur in the   SDW and d-wave SC  channels \cite{Nickel06,Duprat01},  as shown by a singularity in the corresponding  static response functions denoted  by $\chi_{\rm SDW}({\bf q}_0)$ and $\chi_{\rm SCd}(0)$, respectively. In the RG framework,  we have 
\begin{equation}
\label{ }
\chi_\mu({\bf q}_{\mu,0})=(\pi v_F)^{-1}\int_\ell \langle f_\mu(k_b)z^2_\mu(k_b)\rangle_{k_b} d\ell,
\end{equation}
whose $k_b$ average expression  depends on the  vertex renormalisation factors $z_\mu(k_b)$ and the form factors $f_{\rm SDW}=1$ and $f_{\rm SCd}=2\cos^2 k_b$ for the SDW and SCd channels, respectively \cite{Nickel06,Duprat01}.  The $z_\mu'$s obey the flow equation
\begin{equation}
\label{zmu}
\partial_\ell z_\mu(k_b) = \langle (\partial_\ell{\cal L}_\mu ) \tilde{g}_\mu z_\mu({k}'_b)\rangle_{{k}'_b},
\end{equation}
which is governed by the combinations of  momentum dependent couplings $\tilde{g}_{\rm SDW} =\tilde{g}_{2}(k_{b},{k}'_{b},k_{b}+ \pi) + \tilde{g}_{3}(k_{b},k'_{b} + \pi,k_{b} + \pi)$ and $\tilde{g}_{\rm SCd} =-\tilde{g}_1(-k'_{b},k_{b},-k_{b}) -\tilde{g}_2(-k'_{b},-k_{b},k_{b})$. A singularity in $\tilde{g}_{\rm SDW}\ ( {\rm resp.,\ } \tilde{g}_{\rm SCd})$ is  synonymous with a singularity in $\chi_{\rm SDW}\  ({\rm resp.,\ } \chi_{\rm SCd})$ at the ordering temperature $T_{\rm SDW}$  ({\rm resp.,\ } $T_{\rm SCd}$). 

\subsection{Phase diagram, pairing and  spin correlations}
\begin{figure}
 \centerline{ \includegraphics[width=15.0cm]{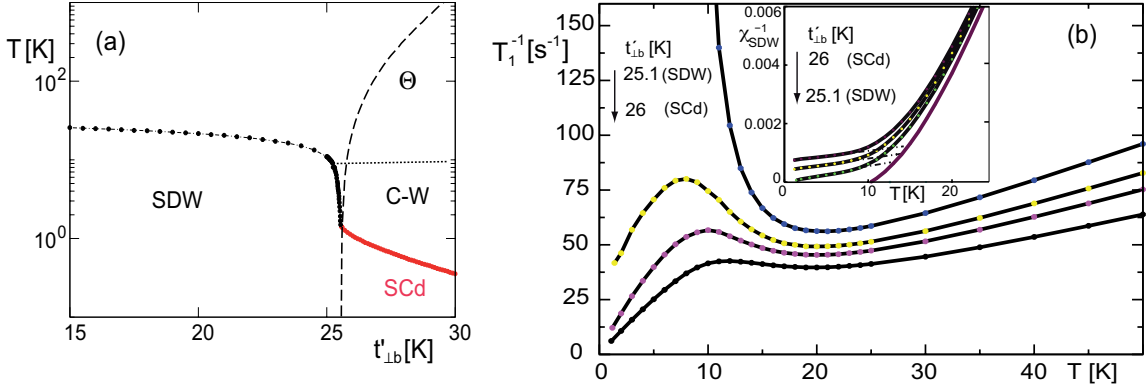}}
\caption{(a) One loop RG phase diagram of the quasi-one-dimensional electron gas model versus the `pressure'  antinesting parameter $t_{\perp b}'$; also shown the variation of the Curie-Weiss (C-W) temperature scale $\Theta$ and the C-W emperature domain. (b) Calculated nuclear spin-lattice relaxation rate $T_1^{-1}$ versus temperature for different $t'_{\perp b}$. Inset: temperature dependence of the inverse SDW susceptibility $(\pi v_F \chi_{\rm SDW})^{-1}$ versus temperature showing the C-W  behavior at low temperature (dashed lines). After Ref.~\cite{Bourbon09}.  \label{PhasesRG}} 
   \end{figure}

From the solution of Eqs. (\ref{gn}) and (\ref{zmu}) and the initial conditions of the model,   one can follow the evolution of the ordering   scale $T_\mu$ as a function of the antinesting parameter $t_{\perp b}'$   simulating the influence of pressure. In the calculated  phase diagram of Fig.~\ref{PhasesRG}-a, we observe that at small $t_{\perp b}'$, nesting is weakly affected  and an instability toward the formation of a SDW state follows at the wave vector ${\bf q}_0=(2k_F,\pi)$ and temperature  \Ts$ \sim 20$K, which is of the order found in systems like (TMTSF)$_2X$ in normal pressure conditions \cite{Jerome82,Bourbon08,Klemme95}. It should be stressed, here, that the above one-loop  calculation yields   a \Ts \ that is strongly reduced    compared to  the  mean-field  limit for which    only  the singularity of the  Peierls scattering channel is retained. This reduction is caused by the influence of the   Cooper channel in (\ref{gn}), whose primary interfering effect is to prevent degrees of freedom in  the 1D energy range  ($E_Fe^{-\ell} > t_{\perp b}$) to produce any instability of the electron gas, in line with the absence of long-range order in one dimension.

As $t_{\perp b}'$ is raised, \Ts\ first smoothly decreases until nesting deterioration approaches  the threshold    $  t'^*_{\perp b}$ where \Ts\   falls sharply (Fig.~\ref{PhasesRG}-a). However, the \Ts\  instability line, instead of dropping toward zero  -- in which case it would lead to a quantum critical point  --   terminates  at  the beginning of a different transition line,  $T_c$, marked  by a   singularity in  the SCd response $\chi_{\rm SCd}$ of the superconducting channel.  The quantum critical point  is therefore avoided by the occurrence of superconductivity. As for the amplitude of $T_c$, it  is maximum  at the juncture    with the SDW phase, giving a ratio $T^0_{\rm SDW}/T^0_c   \sim 20$ between the optimal $T_{\rm SDW}^0$ at zero    $t'_{\perp b}$ and the maximum $T_c$  at $  t'^*_{\perp b}$. Combining these results with those obtained beyond $  t'^*_{\perp b}$  where  $T_c$ is steadily decreasing,  yields   the phase diagram of Figure~\ref{PhasesRG}-a. Singular attractive Cooper pairing is here fed  by  strong SDW  correlations through the interference present in Eqs.~\ref{gn}. The d-wave character of the superconducting transition marks the onset of a SCd order parameter $\Delta(k_{\perp b}) = \Delta \cos k_{\perp b} + {\cal O}(\cos 2k_{\perp b})$, which  indicates that singlet  Cooper pairs  are essentially formed   by electrons of neighboring chains   \cite{Emery86,Caron86B}. Identifying in first approximation the antinesting parameter  $t_{\perp b}'$ as being proportional to real pressure, the essential features of the calculated RG phase diagram of Fig.~\ref{PhasesRG}-a compare  favorably  with the  sequence of transitions  found on experimental grounds (Fig.~\ref{Phases}-a) \cite{Jerome82,DoironLeyraud09}.

The impact  of  interfering SDW-SCd correlations   is not  limited to the sequence of    instabilities lines, but its influence extends also   in the normal phase. This is particularly noticeable when one considers the  temperature dependence of  SDW susceptibility in the superconducting sector of the phase diagram \cite{Bourbon09}. Insert of  Fig.~\ref{PhasesRG}-b shows for example that $\chi_{\rm SDW}({\bf q}_0)$ still increases at the approach of $T_c$, showing the Curie-Weiss  enhancement $\chi_{\rm SDW}({\bf q}_0)\propto (T+\Theta)^{-1}$ below 10~K or so. The origin of the enhancement comes from the positive feedback of SCd Cooper pairing on SDW correlations, which reinforces  staggered spin fluctuations over a Curie-Weiss temperature interval of  about ten times $T_c$. The Curie-Weiss scale $\Theta$  for SDW fluctuations  displays a particular variation with  `pressure' $t_{\perp b}'$: it starts from zero at $  t'^*_{\perp b}$ and then raises rapidly as $ t_{\perp b}'>  t'^*_{\perp b}$, showing that SDW fluctuations decreases in amplitude, which  in turn correlated with the decline of $T_c$.  Considering the point $t'^*_{\perp b}$ at zero temperature as  quantum critical in the absence of superconductivity,  $\Theta$ can   be seen as a characteristic energy scale for SDW fluctuations that increases to some power of the tuning parameter $ t_{\perp b}'$. In the usual terminology of quantum phase transitions, one can write  $\Theta \sim \xi^{-z}$, where $\xi\sim ( t_{\perp b}'-  t'^*_{\perp b})^{-\nu}$ is the zero temperature SDW correlation length close to $t'^*_{\perp b}$. In the one-loop approximation, the exponents are   $\nu=1/2$ for the correlation length and $z=2$ for the dynamics \cite{Bourbon09}, which leads to  a linear  variation of $\Theta$ with $t_{\perp b}'$ in the vicinity  of $ t'^*_{\perp b}$. 
\subsection{Nuclear relaxation rate }
The Curie Weiss behavior of $\chi_{\rm SDW}$ has  also an effect on the temperature dependence of other quantities   sensitive to staggered spin fluctuations. This is  the case of  the  NMR spin-lattice relaxation rate $T_1^{-1}$, whose expression can be written as  an integral of the imaginary part of the spin response over  all  wave vectors ${\bf q}$,
 \begin{equation}
T_1^{-1} = T \int |A_{\bf q}|^2 \ {\chi''({\bf q},\omega)\over\omega}  d^dq.
\label{T1a}
\end{equation}
 The  $T_1^{-1}$ calculation   using the above one-loop RG results   \cite{Bourbon09},  leads to   
 \begin{equation}
T_1^{-1}\approx c_0T + c_1T{\chi_{\rm SDW}({\bf q}_0)\over  \sqrt{1+\xi_c^2}},
\label{T1b}
\end{equation}
where  $\xi_c (\propto \sqrt{\chi_{\rm SDW}})$ is the SDW correlation length along the third direction; $c_0$ and $c_1$ are positive constants.  The expression for $T_1^{-1}$  then superimposes   a linear-$T$ Korringa like  contribution coming from uniform  (${\bf q}\sim 0$) spin fluctuations, which dominates at large temperature, and an anomalous  one coming from SDW $({\bf q}\sim {\bf q}_0$)  fluctuations, which eventually  takes over   at lower  temperature.  In the SDW sector of the  phase diagram where both $\chi_{\rm SDW}$ and $\xi_c$ become singular at $T_{\rm SDW}$, $T_1^{-1} \propto (T-T_{\rm SDW})^{-{1\over 2}}$ develops a square root singularity (Fig.~\ref{PhasesRG}-b), a characteristic of a 3D critical behavior  known to be consonant with   the   experimental situation of Fig.~\ref{T1} \cite{Brown08,Wzietek93}. As one reaches the superconducting region, from $t'^*_{\perp b}$ upwards in  the phase diagram,  $T_1^{-1}$, albeit non longer singular, is strongly enhanced signaling  marked deviations from the  linear-$T$ Korringa  behavior predicted for a Fermi liquid. The anomaly extends deeply in the 2D metallic phase ($\xi_c \ll 1$),  up to at least $20T_c$.  According to (\ref{T1b}),  the amplitude of the enhancement is   correlated   to the one of SDW fluctuations and in turn to the size of $T_c$ as a function of $t'_{\perp b}$.  The Curie-Weiss behavior of $\chi_{\rm SDW}$  is  found  for $(T_1T)^{-1} \sim (T+\Theta)^{-1}$ as well, so that the scale $\Theta$ extracted from the $T_1^{-1}$ data (Fig.~\ref{T1}-b) \cite{Brown08,Wu05,Creuzet87b} coincides  with the one governing the behavior of $\chi_{\rm SDW}$ in the present theory. The interference induced reinforcement of SDW correlations by SCd Cooper pairing thus appears as the mechanism responsible  for the Curie-Weiss behavior observed in  $T_1^{-1}$ \cite{Bourbon09}.

\subsection{Quasi-particle scattering rate and resistivity} 
\begin{figure}
 \centerline{ \includegraphics[width=7.0cm]{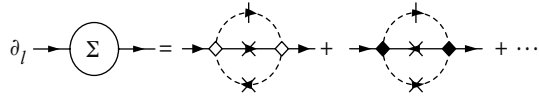}}
\caption{Two-loop RG flow equation for the one-particle self-energy $\Sigma_+$. \label{self}} 
   \end{figure}

 Carrying out the RG transformation up to the two-loop level introduces a renormalization  of the inverse one-particle propagator   $[G_p^0]^{-1}  \to [G_p^0]^{-1} - \Sigma_p$, in the form of Matsubara self-energy  corrections  $\Sigma_p$ (Fig.~\ref{self}). On the Fermi surface ${\bf k}_F(k_b)$, the self-energy takes the form 
 \begin{equation}
\Sigma_p(\bar{k}_b) =   \  i\omega_n [1-z(\bar{k}_b)] 
  - z_\perp(\bar{k}_b)
\end{equation}
 where $z$ and $z_\perp$ are renormalization factors   evaluated at $\bar{k}_b=( {\bf k}_F(k_b),\omega_n)$. The  corresponding flow equations obtained in Ref.~\cite{Sedeki11}, can be written   schematically as  
 \begin{eqnarray}
\partial_\ell \ln z(\bar{k}_b) &= & \sum_{i=1}^3 \sum_{n,n'} \langle \tilde{g}_n\cdot\tilde{g}_{n'} \cdot \partial_\ell {\cal T}_i\rangle_{k_b',q_b'},\\
\partial_\ell   z_\perp(\bar{k}_b) &= &z^{-1}(\bar{k}_b) \sum_{i=1}^3 \sum_{n,n'} \langle \tilde{g}_n\cdot\tilde{g}_{n'} \cdot \partial_\ell {\cal T}_{\perp,i}\rangle_{k_b',q_b'}, 
\end{eqnarray}
where  ${\cal T}_{i,(\perp)}  = \pm T^2 \sum_{\{\omega\}} \int_{{\cal D}_{i,(\perp)}} dE_p [G^0_p]_\ell[G^0_{-p}]_\times [G^0_{-p}]_\times $ corresponds to the expressions of the two-loop diagrams of Figure~\ref{self},  whose sign is  fixed by their topology.   
Solving the   equations for $z$ and $z_\perp$, consistently  with those at the one-loop level for the $\tilde{g}_n$, yields  $\Sigma_p(\bar{k}_b)$  on the Fermi surface for all $\omega_n$. Using a Pad\'e procedure, the Matsubara self-energy  can be  analytically  continued  to the  retarded form of self-energy, which consists of a real [$\Sigma_p'({\bf k}_F(k_b),\omega)]$ and imaginary  [$\Sigma_p''({\bf k}_F(k_b),\omega)]$ parts at real frequency $\omega$. These two  ingredients enter   in the determination of  spectral properties of the quasi-1D electron gas model \cite{Sedeki11}.

\begin{figure}
 \centerline{ \includegraphics[width=15.0cm]{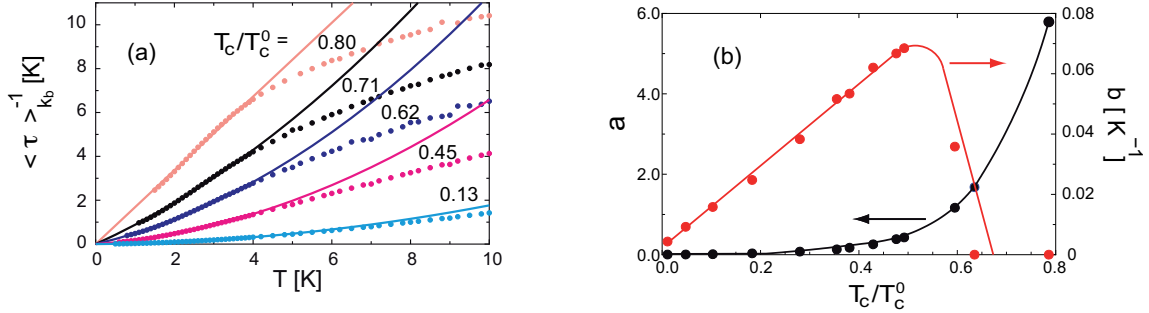}}
\caption{(a) Averaged electron-electron scattering rate versus temperature calculated for different superconducting $T_c$ of the  RG phase diagram (Fig.~\ref{1loop}-a). The continuous lines correspond  to the polynomial fit $\langle \tau\rangle^{-1}_{k_b} = aT + bT^2$ below 4~K. (b) Variation of the  coefficients for a  linear$-T$ and Fermi liquid terms  of the polynomial fit versus $T_c$.  \label{RhoTh}} 
   \end{figure}

 Among quantities  of interest   that can be extracted from the imaginary part  is the electron-electron scattering rate on the Fermi surface 
  \begin{equation}
\tau^{-1}({\bf k}_F(k_b))= -2\Sigma''({\bf k}_F(k_b),\omega\to 0).
\end{equation}
 In   the relaxation time  approximation for conductivity, the average scattering rate  $\langle \tau\rangle^{-1}_{k_b}$ for  an open, strongly anisotropic,  Fermi surface will give the contribution of the singular scattering channels to  the temperature dependence of    electrical resistivity. By excluding the impurity contribution, the latter reads
 \begin{equation}
  \Delta \rho = {4\pi\over \omega_p^2} \langle \tau\rangle^{-1}_{k_b}, 
\end{equation}
  where $\omega_p$ is the plasma frequency, e.g., along the chains. 
The calculated low temperature dependence of $\langle \tau\rangle^{-1}_{k_b}$   down to $T_c$ is shown in Fig.~\ref{RhoTh}-a, as a function of  $T_c(t_{\perp b}')$ in the superconducting region of the phase diagram of Fig.~\ref{PhasesRG}-a. Close to the critical point $t'^*_{\perp b}$, where $T_c$  approaches its maximum, the  behavior of $\langle \tau\rangle^{-1}_{k_b}$, though metallic, does not follow  the $T^2$ law expected for a Fermi liquid above $T_c$. A strict linear behavior,    $\langle \tau\rangle^{-1}_{k_b}= aT$, is instead seen  up to four times $T_c$ or so, namely within the  temperature range where the Curie-Weiss behavior takes  place for $\chi_{\rm SDW}$ and $T_1^{-1}$. We have seen that by moving toward $t'^*_{\perp b}$ from above,   SDW correlations,  stimulated by Cooper pairing, grow in amplitude  ; the  characteristic energy scale of spin fluctuations then decreases to ultimately attain  thermal energy. In two dimensional systems, these conditions are particularly favorable for  the emergence of a linear $T$-behavior in the scattering  rate \cite{Abanov03}, which the above two-loop RG calculations confirm.  By treating density-wave and Cooper pairings on the same footing, the RG also provides an explanation of the origin of the linear $T$-resistivity term seen in compounds like the Bechgaard salts (Fig.~\ref{Phases}-b), and  shows how  interfering orders connect the  amplitude $a$ to  $t_{\perp b}'$ and in turn to the size of $T_c$.  

Moving    away from the SDW-SCd juncture  along the $t_{\perp b}'$ axis, the $\langle \tau\rangle^{-1}_{k_b}$    temperature dependence starts to display some upward curvature [Fig.~\ref{RhoTh}].  It turns out that within a temperature interval of at least four times $T_c$,  the   overall temperature dependence is adequately  fitted    by the polynomial expression 
\begin{equation}
\langle \tau\rangle^{-1}_{k_b} \approx aT + bT^2,
\end{equation}
which can be interpreted as the superimposition of   two scattering channels for electrons under `pressure'. The onset of a Fermi liquid component in the scattering rate results from the continuing stiffening of spin fluctuations, whose  spectral weight moves  to higher frequencies, as shown for example by the evolution of $\Theta$ in nuclear relaxation rate  under `pressure' (Fig.~\ref{PhasesRG}-b). The scattering of electrons on higher energetic modes compared to temperature favors the onset of a $T^2$ component in $ \langle \tau\rangle^{-1}_{k_b} $ \cite{Abanov03,Vilk97}.   
In Fig.~\ref{RhoTh}-b,   the  variation  of the linear and quadratic coefficients $a$ and $b$ is shown  as a function of the ratio $T_c/T_c^0$  for Cooper pairing  tied by interference to the strength  of SDW correlations.  When $T_c/T_c^0 \lesssim 1$, namely  for $t_{\perp b}'\gtrsim t'^*_{\perp b}$, $b$ is vanishingly small and the behavior of the scattering rate  is essentially linear in temperature. However, as    $T_c$ decreases,   $a$ falls approximately as $a\sim T_c^2$, while   $b$ initiates a steep growth    reaching a maximum at intermediate $T_c/T_c^0$ to finally decrease  substantially in the limit of small $T_c$. Such a behavior for $b$ {\it vs} $T_c$ results from the  balance between  the rapid increase of the characteristic energy scale  of SDW fluctuations (compared to temperature) on  the one hand, and the decrease in the amplitude of the same  fluctuations on the  other [Fig.~\ref{PhasesRG}].  The former   is responsible for  the presence  of a Fermi liquid component, whereas the latter governs the decline of both the Fermi liquid and linear contributions over the entire range of  $t_{\perp b}'$ where $T_c$ takes  appreciable  values.  The fact that Cooper pairing  reinforces SDW correlations  by interference, makes existence of a linear component  not limited to the very close proximity of $ t'^*_{\perp b}$, as it would be expected for a quantum critical point, but  is stretched out  over the whole interval where  $T_c$ is discernible.       

The   features depicted   by $a$ and $b$  in Fig.~\ref{RhoTh}-b      find a favorable echo  in electrical  resistivity  of   (TMTSF)$_2$PF$_6$  under pressure (Figs.~\ref{Phases}-b and ~\ref{T1}-a). The results also supply a direct   connection between scattering and spin fluctuations on one hand,  and  Cooper pairing on the other.


\section{Summary and conclusion}
Antiferromagnetism bordering on superconductivity   certainly stands out as the most representative feature of the phase diagram of the Bechgaard salts series of organic conductors under pressure. Far from being unrelated the two types of   orders prove  to be intimately connected, a connection  shown to be   suitably  formulated by the weak  coupling renormalization group  theory of the repulsive quasi-one-dimensional electron gas model.    At the heart of this formulation resides an accurate integration of the two interfering nesting and Cooper pairing mechanisms,  which  condition the instability of the metallic state towards the onset of long-range order. By tuning the amplitude of nesting alterations, interference controls the sequence of spin-density-wave and d-wave superconducting instabilities, a structure that  matches fairly well the one displayed by the phase diagram of the  (TMTSF)$_2$$X$.  Mixed pairings not only exert  influence on   instability lines, but    also affect   short-range spin correlations  whose amplitude is  strengthened by Cooper pairing in the normal state.  Mutual pairing strengthening results in the Curie-Weiss  enhancement of staggered spin fluctuations  which governs, for instance the temperature dependence of the  nuclear spin-lattice relaxation rate   above $T_c$, in agreement with    experimental situation in   the  Bechgaard salts.   The mark left by spin fluctuations  can be found in other quantities  like the   electron-electron scattering rate which is relevant for  the temperature dependence of electrical resistivity. It was shown that it is within the Curie-Weiss domain of spin correlations  that a  linear$-T$ dependence for the scattering rate takes place. Its amplitude correlates with  the strength of Cooper pairing or $T_c$, in line with the evolution of  linear$-T$ resistivity seen  under pressure in the Bechgaard salts.     

The   adherence of the above ideas  with experimental facts indicates that the problem raised by magnetism and Cooper pairing in unconventional superconductors  like the Bechgaard salts can be approached in a unified fashion by means of a weak coupling scaling theory. From a broader viewpoint, this  brings us to     ask in conclusion   whether the same mechanisms are at play in other series of higher $T_c$ unconventional superconductors for which  striking similarities with   the Bechgaard salts  can be found.

\bigskip
The authors acknowledges fruitful and continuous collaboration with  P. Auban-Senzier, D. Bergeron, N. Doiron-Leyraud, D. J\'erome, S. Ren\'e de Cotret and L. Taillefer.  C. B. thanks S. E. Brown for many useful discussions on the topics developed in this work. This work has received the financial support from the National Science and Engineering Research Council  of Canada (NSERC), R\'eseau Qu\'ebcois des Mat\'eriaux de Pointe (RQMP) and  the {\it Quantum materials} program of Canadian Institute of Advanced Research (CIFAR). The authors are also thankful to the R\'eseau Qu\'eb\'ecois de Calcul Haute Performance (RQCHP) for supercomputer facilities at the Universit\'e de Sherbrooke.





\bibliographystyle{model1-num-names}
 \bibliography{/Users/cbourbon/Dossiers/articles/Bibliographie/articlesII.bib} 







\end{document}